\documentclass[prl,twocolumn,showpacs,superscriptaddress]{revtex4}
\usepackage[latin1]{inputenc}
\usepackage{graphicx}
\usepackage{amssymb}
\usepackage{amsmath}
\usepackage{xspace}
\usepackage{bm}
\usepackage{times}

\newfont{\tensy}{cmsy10}

\newcommand{\ie}[0]{i.e.\@\xspace}
\newcommand{\eg}[0]{e.g.\@\xspace}

\newcommand{\UP}[0]{\uparrow}
\newcommand{\DO}[0]{\downarrow}
\newcommand{\nag}{{\phantom{\dag}}}

\newcommand{\las}[0]{\langle}
\newcommand{\ras}[0]{\rangle}
\newcommand{\la}[0]{\left\las}
\newcommand{\ra}[0]{\right\ras}
\newcommand{\ket}[1]{\left|#1\ra}
\newcommand{\bra}[1]{\la#1\right|}

\renewcommand{\tilde}[1]{\widetilde{#1}}

\newcommand{\Tr}[0]{\mbox{Tr}}

\begin{document}

\title{Quantum Fluctuations, Temperature and Detuning Effects in Solid-Light Systems}

\author{Markus Aichhorn}
\affiliation{Institute for Theoretical Physics and Astrophysics, University
  of W\"urzburg, Germany}
\author{Martin Hohenadler}
\affiliation{%
Theory of Condensed Matter, Cavendish Laboratory, University of Cambridge,
Cambridge CB3 0HE, United Kingdom}
\author{Charles Tahan}
\affiliation{%
Theory of Condensed Matter, Cavendish Laboratory, University of Cambridge,
Cambridge CB3 0HE, United Kingdom}
\affiliation{%
Booz Allen Hamilton Inc., 3811 N. Fairfax Dr., Arlington, VA 22203, USA}
\author{Peter B.~Littlewood}
\affiliation{%
Theory of Condensed Matter, Cavendish Laboratory, University of Cambridge,
Cambridge CB3 0HE, United Kingdom}

\begin{abstract}
  The superfluid to Mott insulator transition in cavity polariton arrays is
  analyzed using the variational cluster approach, taking into account
  quantum fluctuations exactly on finite length scales. Phase diagrams in one
  and two dimensions exhibit important non-mean-field features.
  Single-particle excitation spectra in the Mott phase are dominated by
  particle and hole bands separated by a Mott gap. In contrast to
  Bose-Hubbard models, detuning allows for changing the nature of the bosonic
  particles from quasilocalized excitons to polaritons to weakly interacting
  photons. The Mott state with density one exists up to temperatures
  $T/g\gtrsim0.03$, implying experimentally accessible temperatures for
  realistic cavity couplings $g$.
\end{abstract} 

\pacs{71.36.+c, 73.43.Nq, 78.20.Bh, 42.50.Ct} 

\maketitle

The prospect of realizing a tunable, strongly correlated system of photons is
exciting, both as a testbed for quantum many-body dynamics and for the
potential of quantum simulators and other advanced quantum devices. Three
proposals based on cavity-QED arrays have recently shown how this might be
accomplished \cite{GrTaCoHo06,HaBrPl06,AnSaBo07}, followed by further work
\cite{hartmann,Na.Ut.Ti.Ya.07,Le.Le.07,Ma.Co.Ta.Ho.Gr.07,Ro.Fa.07}.
Engineered strong photon-photon interactions and hopping between cavities
allow photons (as a component of cavity polaritons) to behave much like
electrons or atoms in a many-body context. It is clear that a particular
signature of quantum many-body physics, the superfluid (SF) to Mott insulator
(MI) transition, should be reproducible in such systems and be similar to the
widely studied Bose-Hubbard model (BHM). Yet, the BH analogy is not complete.
The mixed matter-light nature of the system brings new physics yet to be
fully explored.

``Solid-light'' systems---so-named for the intriguing MI state of photons
they exhibit---are reminiscent of cold atom optical lattices (CAOL)
\cite{Le.Sa.Ah.Da.Se.Se.07}, but have some advantages concerning direct
addressing of individual sites and device integration, and the potential for
asymmetry construction by individual tuning, local variation, and far from
equilibrium devices \cite{hartmann}. Photons as part of the system serve as
excellent experimental probes, and have excellent ``flying'' potential so
that they can be transported over long distances. Temporal and spatial
correlation functions are accessible, and non-equilibrium quantum dynamics
may be studied using coherent laser pumping to create initial states.  The
possible implementations are many \cite{GrTaCoHo06}. In particular,
microcavities linked by optical fibers \cite{HaBrPl06,Ch.Gr.Mo.Vu.Lu.De.07},
small arrays of stripline superconducting Cooper-pair boxes or ``transmon''
cavities \cite{koch:042319,GrTaCoHo06}, condensate arrays
\cite{La.Ki.Ut.Ro.De.Fr.By.Re.Ku.Fu.Ya.07} and color center/quantum dot
periodic band-gap (PBG) materials \cite{GrTaCoHo06,Na.Ut.Ti.Ya.07} seem most
promising.

Here we focus on the simplest solid-light model \cite{GrTaCoHo06}, describing
$L$ optical microcavities each containing a single two-level atom with states
$\ket{\DO}$, $\ket{\UP}$ separated by energy $\epsilon$. The Hamiltonian
reads ($\hbar=1$)
\begin{eqnarray}\label{eq:ham}
  \hat{H}
  &=&
  -t\sum_{\las i,j\ras} a^\dag_i a^\nag_j + \sum_i \hat{H}^{\text{JC}}_i
  - \mu \hat{N}_\text{p}
  \,,
  \\
  \hat{H}^{\text{JC}}_i
  &=&
  \epsilon \ket{\UP_i}\bra{\UP_i}
  +
  \omega_0 a^\dag_i a^\nag_i
  +
  g (\ket{\UP_i}\bra{\DO_i} a^\nag_i + \ket{\DO_i}\bra{\UP_i} a^\dag_i)
  \,.\nonumber
\end{eqnarray}
Here $\omega_0$ is the cavity photon energy, and $\Delta=\omega_0-\epsilon$
defines the detuning. Each cavity is described by the well-known
Jaynes-Cummings (JC) Hamiltonian $\hat{H}^\text{JC}$. The atom-photon
coupling $g$ ($a^\dag_i$, $a^\nag_i$ are photon creation and annihilation
operators) gives rise to formation of polaritons (combined atom-photon
excitations) whose number $\hat{N}_\text{p}=\sum_i (a^\dag_i a^\nag_i +
\ket{\UP_i}\bra{\UP_i})$ is conserved and couples to the chemical potential
$\mu$ \cite{Ma.Co.Ta.Ho.Gr.07}. We consider nearest-neighbor photon hopping
with amplitude $t$, define the polariton density $n=\las N_\text{p}\ras/L$,
use $g$ as the unit of energy and set $\omega_0/g$ \footnote{As in
  \cite{GrTaCoHo06}; $\omega_0$ is not an absolute frequency, so that this
  choice does not affect results.}, $k_B$ and the lattice constant to one.

Hamiltonian~(\ref{eq:ham}) represents a generic model of strongly correlated
photons amenable to numerical methods. Existing theoretical work has focused
on mean-field calculations \cite{GrTaCoHo06}, exact diagonalization of few
cavity systems \cite{HaBrPl06,AnSaBo07,Ma.Co.Ta.Ho.Gr.07}, and the
one-dimensional case \cite{Ro.Fa.07}.  Here we employ a quantum many-body
method for the thermodynamic limit to explore the physics of the model. In
particular, quantum fluctuations on a finite length scale are included. We
discuss Mott lobes (also at experimentally relevant finite temperatures), the
effect of detuning and for the first time in such systems calculate
single-particle spectra, a necessary connection to experiment and also a key
metric in early proof of concept calculations of CAOL systems.

The Variational Cluster Approach (VCA)---introduced first for strongly
correlated electrons~\cite{po.ai.03}---has previously been applied to the
BHM~\cite{ko.du.06}. The main idea is to approximate the self-energy
$\bm{\Sigma}$ of the infinite system by that of a finite reference system.
The matrix notation includes orbitals and bosonic Matsubara frequencies,
$\bm{\Sigma} = \Sigma_{\alpha \beta}(i\omega_n)$. The optimal choice for
$\bm{\Sigma}$ follows from a general variational principle $\delta
\Omega[\bm{\Sigma}]=0$, $\Omega$ being the grand potential.  Trial
self-energies from isolated clusters (with $L_\text{c}$ sites) are
parametrized by the one-particle parameters
$\bm{\xi}^c=\{t^\text{c},\epsilon^\text{c},\omega_0^\text{c},\mu^\text{c}\}$
of the reference-system Hamiltonian, \ie $\bm{\Sigma} =
\bm{\Sigma}(\bm{\xi}^c)$. For bosons \cite{ko.du.06},
\begin{equation}\label{eq:Omega}
  \Omega = \Omega^\text{c} + \Tr \ln\left({\bm{G}_0}^{-1}-{\bm{\Sigma}}\right)^{-1} 
    -\Tr \ln\left({\bm{G}^\text{c}}\right)\,. 
\end{equation}
Here, $\Omega^\text{c}$, $\bm{G^\text{c}}$, and $\bm{\Sigma}$ are the grand
potential, Green's function, and self energy of an isolated cluster, and
$\bm{G}_0$ is the non-interacting ($g=0$) Green's function. $\bm{G}_0$,
$\bm{G^\text{c}}$, and $\bm{\Sigma}$ are evaluated at bosonic Matsubara
frequencies $i\omega_n$, and traces include frequency summation. The
stationary solution is given by $\partial\Omega /
\partial\bm{\xi}^\text{c}=0$. Traces can be evaluated exactly using only the
poles of the Green's function but not their weights \cite{po.ai.03,ko.du.06}.
The poles $\omega_m$ of ${\bm{G}_0}^{-1}-{\bm{\Sigma}}$ are obtained from a
bosonic formulation of the $Q$-matrix method~\cite{ai.ar.06.2}. At
temperature $T>0$, the required matrix diagonalization restricts $L_\text{c}$
and $T_\text{max}$.  For simplicity, we restrict ourselves to a single
variational parameter $\xi^\text{c}=\omega_0^\text{c}$.

The full quantum dynamics are taken into account exactly on the length scale
of the cluster $L_\text{c}$, and even for $L_\text{c}=1$ VCA results are
beyond the mean-field solution \cite{ko.du.06,GrTaCoHo06}.  The present
formulation cannot describe the properties of the SF phase, as the required
symmetry-breaking term $\hat{H}_\psi = \psi\sum_i(a^\dagger_i + a^\nag_i)$
cannot be cast into a single-particle operator. However, this does not affect
the accuracy of the phase boundary of the MI.  The numerical effort is very
moderate as compared to, \eg, the density matrix renormalization group
\cite{Ro.Fa.07}, and the VCA provides $T=0$ and $T>0$ static and dynamic
properties in one and two dimensions.

\begin{figure}
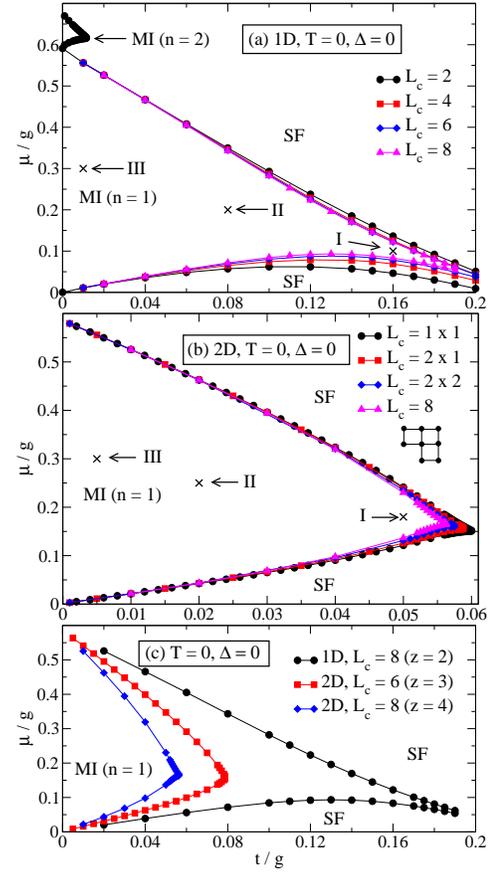

  \centering
  \includegraphics[width=0.35\textwidth]{mottlobes_T0_1D.eps}
  \includegraphics[width=0.35\textwidth]{mottlobes_T0_2D.eps}
  \includegraphics[width=0.35\textwidth]{mottlobes_T0_coord.eps}
  \caption{\label{fig:pdT0}%
    (Color online) Ground-state phase diagram: (a) 1D, (b) 
    2D and (c) different coordination numbers $z$. 
    Lines are guides to the eye.}
\end{figure}

The $T=0$ phase diagram in one (chain) and two dimensions (square lattice) is
shown in Fig.~\ref{fig:pdT0}. There exists a series of Mott lobes with
integer polariton density $n_\text{int}=0,1,\dots$ and compressibility
$\kappa\equiv\partial n/\partial\mu=0$ \cite{GrTaCoHo06}.  In contrast to
recently proposed photonic MI phases \cite{hartmann}, $\las a^\dag a\ras$
fluctuates even for constant, integer $n$. Inside the lobes, where the VCA
yields a solution, the system has an energy gap $\tilde{E}_{g,\text{p}}$
($\tilde{E}_{g,\text{h}}$) for adding a particle (hole) equal to the vertical
distance of $\mu$ from the upper (lower) phase boundary \footnote{For $t/g=0$
  and density $n$, a particle/hole corresponds to one site having occupation
  $n+1$ ($n-1$).}. The spacing of the points
$\mu_n=\omega_0+g(\sqrt{n}-\sqrt{n+1})$ where adjacent lobes touch at $t/g=0$
decreases quickly with increasing $n$, in contrast to the BHM where
$\mu_n=Un$. Convergence with cluster size $L_\text{c}$ is surprisingly fast.
The 1D data agree well with exact results, although the VCA slightly
underestimates $t^*$ (the value of $t$ at the lobe tip, $t^*/g=0.2$ in
\cite{Ro.Fa.07}). Figure~\ref{fig:pdT0}(b) represents the first accurate
(non-mean-field) phase diagram in 2D, arguably the most important case for
experimental realizations.

For $t/g=0$, the MI states are $\prod_i \ket{\psi_n}_i$, where the
$n$-polariton eigenstate of $\hat{H}^\text{JC}$ (the $\ket{-,n}$ branch in
\cite{GrTaCoHo06}) is a superposition of photonic ($\ket{\DO,n}$, with $n$
photons) and excitonic states ($\ket{\UP,n-1}$),
\begin{equation}
  \ket{\psi_n}
  =
  \eta(\Delta,n) \ket{\DO,n}
  +
  \phi(\Delta,n) \ket{\UP,n-1}
  \,.
\end{equation}
Lobes with $n>1$ are much smaller due to the effective polariton repulsion
decreasing with $n$, and we focus on the $n=1$ case for which quantum effects
are strongest, and which can be easily initialized experimentally.

Figures~\ref{fig:pdT0}(a) and (b) show significant deviations from the
parabolic lobes predicted by mean-field theory in both one and two
dimensions. However, quantum fluctuation effects diminish quickly with
increasing coordination number $z$ (Figure~\ref{fig:pdT0}(c)). In particular,
the reentrant behavior with increasing $t/g$ \cite{PhysRevB.58.R14741} and
the cusplike tip indicative of the Berezinskii-Kosterlitz-Thouless transition
\cite{PhysRevB.40.546} exist only for $z=2$.

\begin{figure}
  \centering
  \includegraphics[width=0.45\textwidth]{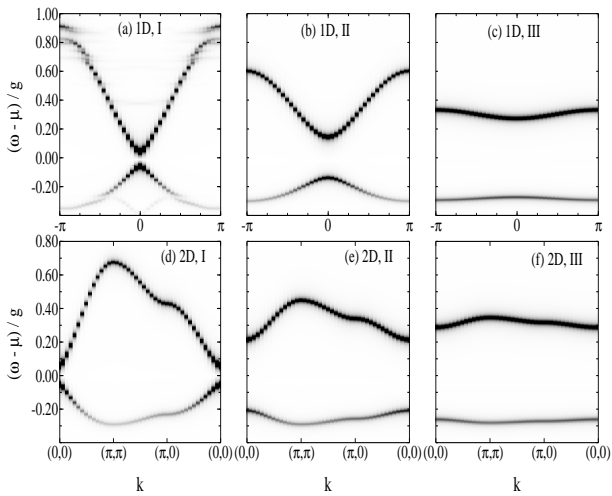}
  \caption{\label{fig:spectra}%
    Single-particle spectra at $T=0$ in 1D [(a)--(c),
    $L_\text{c}=6$] and 2D [(d)--(f),
    $L_\text{c}=2\times2$]. I--III refer to the marks in
    Fig.~\ref{fig:pdT0}, \ie,  
    (a) $t/g=0.16$, (b) 0.08, (c) 0.01, (d) 0.05, (e)
    0.02, (f) 0.005.}   
\end{figure}

Spectral properties play a key role in understanding condensed matter
systems, but are notoriously difficult to calculate accurately. In cavities,
the occupation and spectra can be directly monitored through luminescence
spectra, with angular emission translating into momentum $\bm{k}$, much more
straightforward than in CAOL.

The VCA yields the single-particle spectral function
$A(k,\omega)=-{\pi}^{-1}\text{Im}\,G(k,\omega)$, shown in
Fig.~\ref{fig:spectra} for the parameters marked in Fig.~\ref{fig:pdT0}.
Important for detection in experiment, the MI state is characterized by
cosinelike particle and hole bands, separated by the Mott gap $E_g$, minimal
at $\bm{k}=0$, which decreases with increasing $t/g$, and eventually closes
in the SF phase.  The phase boundaries $\mu_\pm$ are related to the
particle/hole dispersions $\varepsilon_\text{p/h}(\bm{k}=0,t)$ via
$\mu_+(t)=\varepsilon_\text{p}(0,t)$ and
$\mu_-(t)=\varepsilon_\text{h}(0,t)$, and the Mott gap is given by
$E_g(t)=\varepsilon_\text{p}(0,t)-\varepsilon_\text{h}(0,t)$
\cite{PhysRevB.59.12184}.

For $t\ll t^*$, particles are much lighter than holes (\eg,
$tm_\text{p}\approx0.33$, $tm_\text{h}\approx0.81$ in
Fig.~\ref{fig:spectra}(c); obtained from parabolic fits).  The particle/hole
bandwidth scales almost linearly with $t$, confirming the weakly interacting
Bose gas picture for the superfluid fraction of particles/holes doped into
the MI \cite{PhysRevB.40.546}. The hole bandwidth in both one and two
dimensions is about $zt$. For the BHM model, the ratio of the bandwidths is
$W_\text{p}/W_\text{h}=2$ for the MI with $n=1$ due to the effective particle
(hole) hopping $t_\text{p}=(n+1)t$ ($t_\text{h}=nt$)
\cite{PhysRevA.63.053601,ko.du.06}. For the JCM, the matrix elements for
hopping of one particle or hole in the MI have to be evaluated using the
dressed states $\ket{\psi_n}$, and we find
$W_\text{p}/W_\text{h}=3/2+\sqrt{2}\approx2.91$ in one dimension, in good
agreement with Figs.~\ref{fig:spectra}(a)--(c). The bandwidth ratio is fairly
independent of $t/U$ respectively $t/g$ because the interaction energy of a
single particle or hole is the same at every site.  In the BHM, emergent
particle-hole symmetry leads to $m_\text{p}\approx m_\text{h}\to0$ for $t\to
t^*$ \cite{capogrosso-sansone:134302}. This behavior is also seen in the JCM
($m_\text{p}/m_\text{h}\approx1.3$ for $t=0.16$ in 1D), but the VCA does not
permit a detailed analysis near $t^*$.

\begin{figure}
  \centering
  \includegraphics[width=0.35\textwidth]{mottlobes_T0_1D_delta2.eps}
  \includegraphics[width=0.35\textwidth]{mottlobes_T0_1D_delta-2.eps}
  \includegraphics[width=0.35\textwidth]{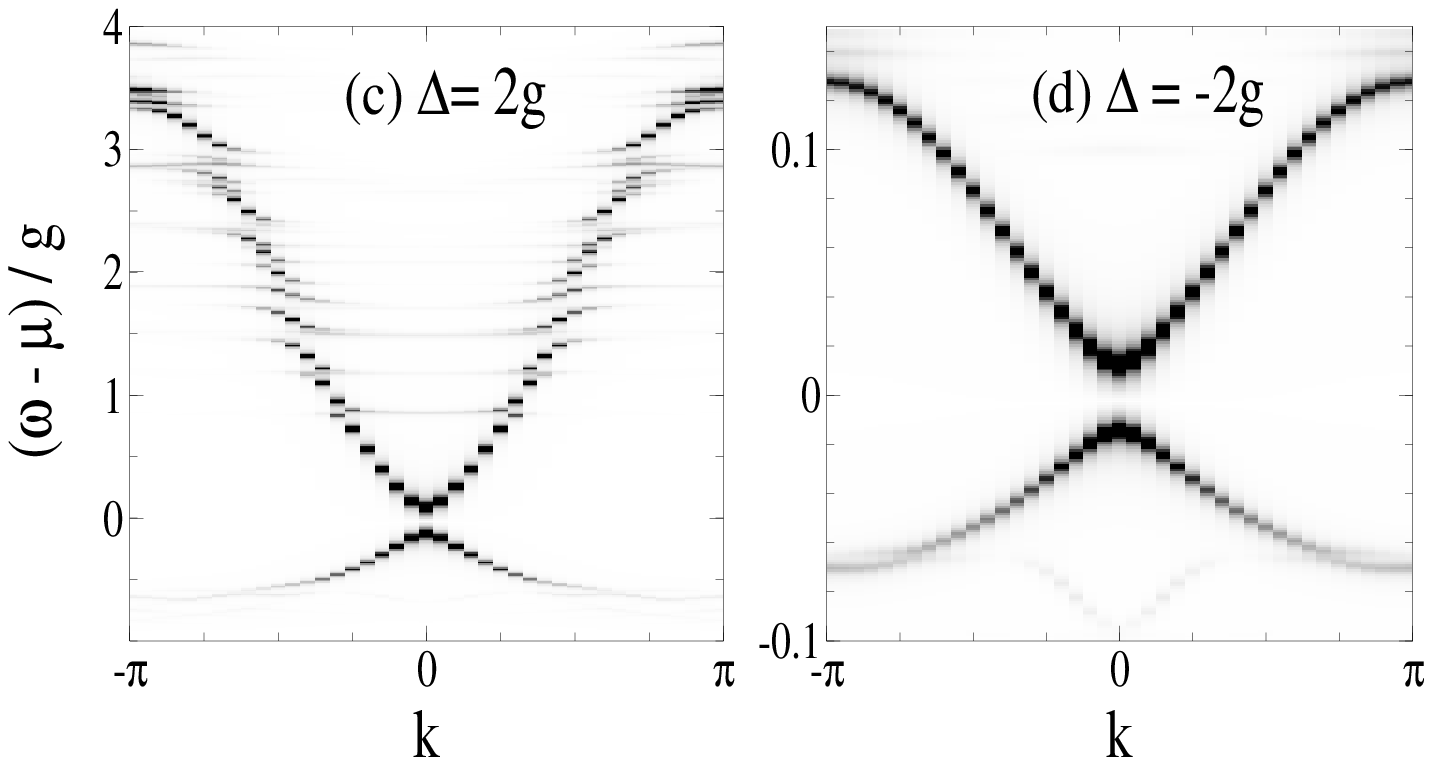}
  \caption{\label{fig:detuned}%
    (Color online) Phase diagram for detuning (a) $\Delta/g=-2$, (b)
    $\Delta/g=2$, and spectra for (c) $\Delta/g=2$, $t/g=0.7$, and (d)
    $\Delta/g=-2$, $t/g=0.0.018$ in 1D at $T=0$. 
}
\end{figure}

In contrast to CAOL, the parameters $\epsilon$ and $\omega_0$ and hence the
detuning $\Delta=\epsilon-\omega_0$ can easily be changed experimentally.
This permits to tune the character of the bosonic particles.  The states
$\ket{\psi_n}$ are excitonlike for $\omega_0\gg\epsilon$, photonlike for
$\omega_0\ll\epsilon$, and polaritonlike for $\omega_0\approx\epsilon$. The
1D phase diagrams for $\Delta/g=\pm2$ are shown in
Fig.~\ref{fig:detuned}(a),(b). The excitonic system ($\Delta/g=2$) with small
photon-mediated hopping exhibits a large $n=1$ Mott lobe, whereas the latter
is very small in the photonic system with small exciton-mediated interaction.
Reentrant behavior due to quantum fluctuations is seen for the excitonic
case, but is absent in the photonic case for large enough $L_\text{c}$.

Single-particle spectra in one dimension near the lobe tips are shown in
Fig.~\ref{fig:detuned}(c),(d). The excitonic system shows a very large ratio
of particle and hole bandwidths ($W_\text{p}/W_\text{h}\approx 7$), whereas
$W_\text{p}/W_\text{h}\approx2.1$ (very close to the BHM, since
$\ket{\psi_1}\approx\ket{\DO,1}$) for the photonic case. These values result
from the different admixture of the states $\ket{\UP,0}$ and $\ket{\DO,1}$ to
$\ket{\psi_1}$ depending on $\Delta$. In particular, the approximate relation
$W_\text{h}\approx zt$ found for $\Delta=0$ does not hold. The excitonic MI
with $n=1$ is approximately given by $\prod_i \ket{\UP,0}_i$, whereas we have
$\prod_i \ket{\DO,1}_i$ for the photonic case. Finally, the incoherent
features in Fig.~\ref{fig:detuned}(c) may originate from the finite cluster
size $L_\text{c}$, and should be addressed using other exact methods.

Solid state quantum devices will operate at finite temperature, much higher
than in CAOL (estimated at nK). This leads to the important question of the
stability of the MI state at $T>0$, which will ultimately determine their
technological usefulness. Strictly speaking, there is no true MI at $T>0$ due
to thermal fluctuations. However, there exist regions where fluctuations are
small enough for the system to behave like a MI for experimental purposes
\cite{PhysRevA.67.033606}. We determine the region of existence of the MI
using the stringent criterion $\Delta n=|n-n_\text{int}|\leq10^{-4}$,
corresponding to a ``worst case scenario'' since experimentally the MI will
survive as long as the Mott gap $\sim g$ is large compared to thermal
fluctuations.

\begin{figure}
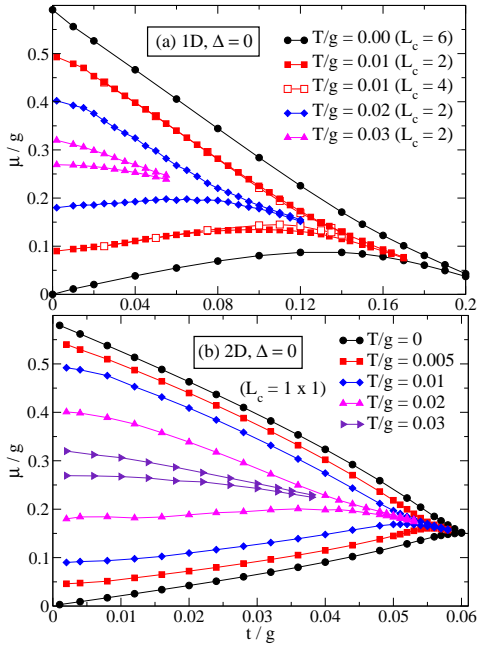

  \centering
  \includegraphics[width=0.35\textwidth]{mottlobes_Tvar_1D.eps}
  \includegraphics[width=0.35\textwidth]{mottlobes_Tvar_2D.eps}
  \caption{\label{fig:pdT>0}%
    (Color online) Mott-like regions for different $T$.}
\end{figure}

Results for $\Delta=0$ are shown for one and two dimensions in
Fig.~\ref{fig:pdT>0}.  Finite-size effects diminish quickly with increasing
temperature (see Fig.~\ref{fig:pdT>0}(a)). The size of the $n=1$ Mott lobe is
significantly reduced with increasing temperature, and $\Delta n$ exceeds
$10^{-4}$ at $T^*/g\approx0.03$ in both one and two dimensions. This value is
consistent with the onset of deviations of $n$ from 1 in the exact
atomic-limit results, and agrees with the few-cavity results of
\cite{Ma.Co.Ta.Ho.Gr.07}. Lobes with $n>1$ disappear at much lower
temperatures (not shown).  In contrast to $T=0$, Mott lobes with $n$ and
$n\pm1$ do not touch at $t/g=0$, but are surrounded by the normal fluid (NF)
phase for sufficiently small $t/g$, and a transition NF-SF occurs at larger
$t/g$. However, while the VCA can distinguish between the MI and the NF, the
phase boundary to the SF at $T>0$ cannot be determined accurately. The gap in
$A(\bm{k},\omega)$ increases with $T$ \cite{PhysRevA.67.033606}.
Particle/hole bands are still well defined at $T>0$, and spectra (not shown)
look very similar to $T=0$.  Excitation spectra of the NF and SF deserve
detailed future studies.

The quasi-MI lobes in Fig.~\ref{fig:pdT>0} are dominated by the point
$(t=0,\mu\approx0.3)$ where $\tilde{E}_g$ is maximal and by the $T=0$ lobe
tip $(\mu^*,t^*)$.  The MI is destroyed by thermal excitation of particles or
holes, and therefore survives longest near $\mu=0.3$ where the energy cost is
largest.  Besides, $T^*$ is determined by the size of the lobe at $T=0$,
$t=0$. Hence, the system at $T>0$ is dominated by atomic-limit physics
(yielding the same $T^*$ in 1D and 2D) and the $T=0$ fixed point (quantum
critical point) of the MI-SF transition \cite{PhysRevB.40.546}.  For the 1D
BHM \cite{PhysRevB.40.546}, the VCA yields $T^*/U\approx0.059$, lower than
$T^*/U=0.2$ found in \cite{gerbier:120405} using a different criterion
\footnote{There are no well-defined plateaus in the density even below
  $T^*/U=0.2$ in \cite{gerbier:120405}.}. The ratio of the critical
temperatures for the 1D JCM and 1D BHM is 0.56, close to the ratio of the
$t=0$, $T=0$ Mott gaps (0.59).

Solid state cavity-QED systems offer the possibility of large $g$.  Taking a
conservative estimate $g=10^{10}$ Hz \cite{GrTaCoHo06} gives for the
effective temperature of the quantum system $T^*\gtrsim14$ mK.  Even taken as
the actual temperature, this is well accessible experimentally. The hopping
rate $1/t$ has to be fast enough to permit equilibration before photon loss
and dephasing set in. Indeed, taking $t/g=0.01$ (inside the Mott lobe in
Fig.~\ref{fig:pdT>0}(b)), we obtain a realistic $t^{-1}\gtrsim0.14$ ps.
Finally, $T^*$ for the $n=1$ MI is enhanced by detuning $\Delta>0$
(Fig.~\ref{fig:detuned}(a)), which additionally increases $t^*$ and
$|\mu^*|$, estimated as $(\mu^*,t^*)\approx(0.08,0.02)$ meV in 2D for
$T=\Delta=0$.

In summary, motivated by theoretical predictions and ongoing experimental
advances, we have studied polariton Mott phases by means of a versatile
quantum many-body approach. Phase diagrams, single-particle spectra and
finite temperature effects have been related to the known features of
Bose-Hubbard models and possible experiments, and the novel physics emerging
from detuning has been explored.

This work was supported by the FWF Schr\"odinger Fellowship No.~J2583, the
DFG research unit FOR538, and by NSF Award No.~DMR-0502047.


\end{document}